\documentclass[preprint,showpacs,preprintnumbers,amsmath,amssymb,superscriptaddress,11pt,graphicx]{revtex4}

\usepackage{graphicx}
\usepackage{dcolumn}
\usepackage{bm}
\usepackage{amssymb}

\def\MNRAS{Mon. Not. Roy. Astron. Soc.}
\newcommand{\be}{\begin{equation}}
\newcommand{\ee}{\end{equation}}
\newcommand{\bea}{\begin{eqnarray}}
\newcommand{\ena}{\end{eqnarray}}
\newcommand{\nn}{\nonumber}

\begin{document}

\title{Determination of Dark Energy by the Einstein Telescope: Comparing with CMB, BAO and SNIa Observations }

\author{W. Zhao}
\affiliation{School of Physics and Astronomy, Cardiff University,
Cardiff, CF24 3AA, United Kingdom }

\author{C. Van Den Broeck}
\affiliation{Nikhef -- National
Institute for Subatomic Physics, Science Park 105, 1098 XG
Amsterdam, The Netherlands }

\author{D. Baskaran}
\affiliation{School of Physics and Astronomy, Cardiff University,
Cardiff, CF24 3AA, United Kingdom }

\author{T.G.F. Li}
\affiliation{Nikhef -- National Institute for Subatomic Physics,
Science Park 105, 1098 XG Amsterdam, The Netherlands }

\date{\today}


\begin{abstract}
{\small A design study is currently in progress for a
third-generation gravitational-wave (GW) detector called Einstein
Telescope (ET). An important kind of source for ET will be the
inspiral and merger of binary neutron stars (BNS) up to $z \sim
2$. If BNS mergers are the progenitors of short-hard $\gamma$-ray
bursts, then some fraction of them will be seen both
electromagnetically and through GW, so that the luminosity
distance and the redshift of the source can be determined
separately. An important property of these `standard sirens' is
that they are \emph{self-calibrating}: the luminosity distance can
be inferred directly from the GW signal, with no need for a cosmic
distance ladder. Thus, standard sirens will provide a powerful
independent check of the $\Lambda$CDM model. In previous work,
estimates were made of how well ET would be able to measure a
subset of the cosmological parameters (such as the dark energy
parameter $w_0$) it will have access to, assuming that the others
had been determined to great accuracy by alternative means. Here
we perform a more careful analysis by explicitly using the
potential Planck cosmic microwave background data as prior
information for these other parameters. We find that ET will be
able to constrain $w_0$ and $w_a$ with accuracies $\Delta w_0 =
0.099$ and $\Delta w_a = 0.302$, respectively. These results are
compared with projected accuracies for the JDEM Baryon Acoustic
Oscillations project and the SNAP Type Ia supernovae observations.

}

\end{abstract}

\pacs{98.70.Vc, 98.80.Cq, 04.30.-w}

\maketitle


\section{Introduction \label{section1}}

In the past decade, various observations, including type-Ia
supernovae (SNIa) \cite{SNIa}, the temperature and polarization
anisotropies power spectrum of the cosmic microwave background
(CMB) radiation \cite{wmap}, the baryon acoustic oscillations
(BAO) peak in the distribution of Sloan Digital Sky Survey
luminous red galaxies \cite{bao}, and weak gravitational lensing
\cite{wl}, have all suggested that the present Universe is
undergoing an accelerated expansion. A possible explanation would
be the presence of a fluid called dark energy, which should have
positive density but negative pressure (for a review, see
\cite{darkenergy}). Understanding the physical character of dark
energy, assuming it exists, is one of the main challenges of
modern cosmology. A key question is then how well we will be able
to differentiate between various dark energy models by measuring
the dark energy equation of state (EOS) and its time evolution.

Currently, among the main methods to determine the dark energy EOS
are observations of SNIa, CMB, and large-scale structure. The
capabilities of these methods will be improved significantly in
the near future \cite{detf,euclid}. However, we note that all
these methods are based on the observations of various
electromagnetic waves. In addition to these electromagnetic
methods, the observation of gravitational waves (GW) will provide
a new technique, where gravitational-wave sources, in particular
inspiraling and merging compact binaries, can be considered as
standard candles, or \emph{standard sirens} \cite{schutz}. In the
case of ground-based detectors, the idea is to use binaries
composed of two neutron stars (BNS), or a neutron star and a black
hole (NSBH). These are hypothesized to be at the origin of
short-hard $\gamma$-ray bursts (shGRBs). In many cases it is
possible to identify the host galaxy of a shGRB and determine its
redshift. From the gravitational-wave signal itself one would be
able to measure the luminosity distance in an absolute way,
without having to rely on a cosmic distance ladder: standard
sirens are \emph{self-calibrating}.

The use of GW standard sirens to measure the Hubble constant with
a network of advanced ground-based detectors has been studied by
Nissanke et al.~\cite{nissanke}, and with LISA (using extreme mass
ratio inspirals) by MacLeod and Hogan \cite{hogan}. Supermassive
binary black holes may be useful to study dark energy with LISA
\cite{lisa1,lisa2,lisa3,lisa4,lisa5,lisa6}; more generally, they
can constrain alternative theories of cosmology and gravity
\cite{DeffayetMenou07,Kocsisetal08,BBW,ArunWill}. Observations of
BNS events with the Big Bang Observer would also allow for dark
energy studies \cite{BBO}.

Currently a third-generation ground-based observatory called
Einstein Telescope (ET) is undergoing a design study \cite{et}.
The latter would be able to see BNS inspirals up to redshifts of
$z \sim 2$ and NSBH events up to $z \sim 8$, corresponding to
millions of sources over the course of several years, some
fraction of which will have a detectable electromagnetic
counterpart (e.g., a shGRB). Sathyaprakash et al.~have
investigated how well cosmological parameters could be determined
with ET assuming 1000 `useful' sources \cite{sath}. Among the
parameters which ET will have access to are \be (H_0, \Omega_m,
\Omega_k, w_0, w_a), \label{parameters} \ee where $H_0$ is the
Hubble parameter at the current epoch, $\Omega_m$ the density of
matter by the critical density, $\Omega_k$ a parameter related to
spatial curvature, and $w_0$ and $w_a$ parameters determining the
dark energy EOS and its time evolution (see below for more precise
definitions). ET will not be able to arrive at a completely
independent measurement of all these parameters at once. In
\cite{sath} it was assumed that, e.g., all parameters except $w_0$
had been measured by other means (electromagnetic or GW) and could
be assumed known with arbitrary accuracy for all practical
purposes. Here we continue this study in more depth, with a focus
on the dark energy parameters $w_0$ and $w_a$. Instead of assuming
the other parameters to be exactly known, we will use the
predicted CMB prior from Planck. CMB measurements give accurate
values for $H_0$, $\Omega_m$, $\Omega_k$, but have large
uncertainties in $w_0$ and $w_a$. Heuristically, imposing this
prior effectively `fixes' the values of $H_0$, $\Omega_m$,
$\Omega_k$. To measure $w_0$, $w_a$ with GW standard sirens is
then an important check of the values obtained through
electromagnetic means.

The outline of the paper is as follows. In Sec.~\ref{section2}, we
recall the basics of using short-hard $\gamma$-ray bursts as
standard sirens in potential ET-GW observations. We then discuss
the determination of the dark energy parameters by the ET-GW
method alone, after which we impose the Planck CMB prior. In
Sec.~\ref{section3}, we discuss the capabilities of the JDEM BAO
project and of the SNAP SNIa project, and a comparison as well as
the potential combination with the ET-GW method is given. In
Sec.~\ref{section4} we conclude with a summary of our main
results.


\section{Short-hard $\gamma$-ray bursts as a kind of standard sirens \label{section2}}

\subsection{The expanding Universe and the dark energy\label{section2a}}

We will work with the Friedmann-Lema\^{i}tre-Robertson-Walker
universes, which are described by:
 \bea
 ds^2=-dt^2+a^2(t)\left\{\frac{dr^2}{1-kr^2}+r^2d\theta^2+r^2\sin^2\theta
 d\phi^2\right\},
 \ena
where $t$ is the cosmic time, and $(r,\theta,\phi)$ are the
comoving spatial coordinates. The parameter $k=0, 1,-1$ describes
the flat, closed and open universe, respectively. The evolution of
the scale factor $a(t)$ depends on the matter and energy contents
of the Universe. Within general relativity, the equations
governing this evolution are
 \bea
 \left(\frac{\dot{a}}{a}\right)^2\equiv H^2=\frac{8\pi
 G\rho_{tot}}{3}-\frac{k}{a^2},~~~~\frac{\ddot{a}}{a}=-\frac{4\pi
 G}{3}(\rho_{tot}+3p_{tot}), \label{friedmann}
 \ena
where $\rho_{tot}$ and $p_{tot}$ are the total energy density and
pressure in the Universe, and $H$ is the Hubble parameter. Since
in this paper we are mainly interested in the later stages of the
evolution of the Universe, where the radiation component can be
ignored, we only take into account baryonic matter, dark matter,
and dark energy. The baryon and dark matter are both modeled as
pressureless dust. We will assume that the EOS of a dark energy
component is responsible for the recent expansion of the Universe,
which should be determined by observations. In this paper, we
shall adopt a phenomenological form for the equation-of-state
parameter $w$ as a function of redshift $z$: \bea
w(z) &\equiv& p_{de}/\rho_{de} = w_0+w_a (1 -a) + \mathcal{O}\left[(1 - a)^2\right] \nn\\
     &\simeq& w_0 + w_a \frac{z}{1 + z} ~.
\label{eos} \ena This is the so-called Chevallier-Polarski-Linder
form \cite{CPL}, which has been adopted by many authors, including
the DETF (Dark Energy Task Force) group \cite{detf}. In the
present epoch where $z \simeq 0$, we have $w \simeq w_0$. $w_a$
describes the evolution of $w$ to next-to-leading order in
$(1-a)$. Since we are mostly interested in the later stages of the
Universe's evolution, higher order terms will be ignored.

The evolution of dark energy is determined by the energy
conservation equation $\dot{\rho}_{de}+3H(\rho_{de}+p_{de})=0$. By
using the EOS of dark energy, Eq.~(\ref{eos}), we find that
 \bea
 \rho_{de}=\rho_{de,0}\times E(z), \label{rho-de}
 \ena
 where $\rho_{de,0}$ is the value of $\rho_{de}$ at $z=0$, and
 \bea
 E(z)\equiv(1+z)^{3(1+w_0+w_a)}e^{-3w_az/(1+z)}.
 \ena
Using Eq.~(\ref{friedmann}), the Hubble parameter $H$ then becomes
 \bea
 H(z)=H_0\left[\Omega_m(1+z)^3+\Omega_k(1+z)^2+(1-\Omega_m-\Omega_k)E(z)\right]^{1/2}.
 \ena
In this expression, $\Omega_m\equiv8\pi G \rho_{m,0}/3H_0^2$ is
the density of matter (baryon as well as dark matter) relative to
the critical density, and $\Omega_k\equiv -k/H_0^2$ is the
contribution of the spatial curvature. $H_0$ is the Hubble
parameter at the present epoch. Throughout this paper, we shall
adopt a fiducial cosmological model with the following values for
the parameters \cite{wmap}: \bea
w_0=-1,~~w_a=0,~~\Omega_bh_0^2=0.02267,~~\Omega_ch_0^2=0.1131,~~
\Omega_k=0,~~h_0=0.705, \ena where $h_0 =
H_0/(100\,\mbox{km}\,\mbox{s}^{-1}\,\mbox{Mpc}^{-1})$. The other
parameters are obtained as $\Omega_m=\Omega_b+\Omega_c=0.2736$,
$\Omega_{de}=1-\Omega_{m}-\Omega_k=0.7264$. In Sec.
\ref{section2d}, the CMB prior for the dark energy determination
will be discussed, where the perturbation parameters $A_s$ and
$n_s$ (the amplitude and spectral index of primordial density
perturbations, respectively) and the reionization parameter $\tau$
(the optical depth of reionization) are also needed. In our
fiducial model, we take these to be \cite{wmap} \bea
A_s=2.445\times10^{-9},~~n_s=0.96,~~\tau=0.084. \ena

To conclude this subsection, we state the expression for the
luminosity distance $d_{L}$ of the astrophysical sources as a
function of redshift $z$ (see, e.g., \cite{detf}): \bea
d_L(z)=(1+z)\left\{
\begin{array}{c}
      {{|k|}^{-1/2}}\sin\left[{|k|}^{1/2}\int^z_0\frac{dz'}{H(z')}\right]~~~(\Omega_k<0),\\
    \int^z_0\frac{dz'}{H(z')}~~~~~~~~~~~~~~~~~~~~~~~~~~~~(\Omega_k=0), \label{dl}\\
      {{|k|}^{-1/2}}\sinh\left[{|k|}^{1/2}\int^z_0\frac{dz'}{H(z')}\right]~~~(\Omega_k>0),
 \end{array}
 \right.
 \ena
where $|k|^{1/2}\equiv H_0\sqrt{|\Omega_k|} $. This formula will
be used frequently in the subsequent discussion.

\subsection{Short-hard $\gamma$-ray bursts and gravitational waves\label{section2b}}

Current observational studies of dark energy strongly rely on
\emph{standard candles}, i.e., sources for which the intrinsic
luminosity is assumed to be known within a certain tolerance so
that they can be used to determine luminosity distance. A widely
used standard candle is the type-Ia supernova (SNIa)
\cite{SNIa1999,SNIa}. The intrinsic luminosity of distant SNIa
needs to be calibrated by comparison with different kinds of
closer-by sources, leading to a `cosmic distance ladder'. This is
not the case with GW standard sirens. As pointed out by Schutz,
the chirping GW signals from inspiraling compact binary stars
(neutron stars and black holes) can provide an absolute measure of
distance, with no dependence on other sources \cite{schutz}. The
GW amplitude depends on the so-called chirp mass (a certain
combination of the component masses) and the luminosity distance.
However, the chirp mass can already be measured from the signal's
phasing, so that the luminosity distance can be extracted from the
amplitude.

Before discussing standard sirens in more detail, let us first
recall some basic facts about the gravitational radiation emitted
by inspiraling compact binaries. Gravitational waves are described
by a second rank tensor $h_{\alpha\beta}$, which, in the so-called
transverse-traceless gauge, has only two independent components
$h_{+}$ and $h_{\times}$, $h_{xx}=-h_{yy}=h_{+}$,
$h_{xy}=h_{yx}=h_{\times}$, all other components being zero. A
detector measures only a certain linear combination of the two
components, called the response $h(t)$, which is given by (see,
e.g., \cite{gwreview}) \bea
h(t)=F_{+}(\theta,\phi,\psi)h_{+}(t)+F_{\times}(\theta,\phi,\psi)h_{\times}(t),
\ena where $F_{+}$ and $F_{\times}$ are the detector antenna
pattern functions, $\psi$ is the polarization angle, and
$(\theta,\phi)$ are angles describing the location of the source
on the sky, relative to the detector. In general these angles are
time-dependent. In the case of Einstein Telescope, binary neutron
star signals can be in band for hours, but almost all of the
signal-to-noise ratio will be accumulated in the final minutes of
the inspiral process. With LISA, Doppler modulation due to the
orbital motion, as well as spin precession, will allow for
accurate determination of the angular parameters (see, e.g.,
\cite{triassintes} and references therein), but this is unlikely
to happen for BNS (or NSBH) signals in ET with Doppler modulation
due to the Earth's rotation. Nevertheless, some improvement in
parameter estimation can be expected, which for simplicity we do
not take into account here. In the sequel, $(\theta,\phi,\psi)$
will be considered constant.


Consider a coalescing binary at a luminosity distance $d_L$, with
component masses $m_1$ and $m_2$. Write $M = m_1 + m_2$ for the
total mass and $\eta = m_1 m_2/M^2$ for the symmetric mass ratio,
and define the `chirp mass' as $\mathcal{M}_c = M \eta^{3/5}$. For
sources at cosmological distances, what enters the waveform is the
\emph{observed} chirp mass, which differs from the \emph{physical}
chirp mass by a factor $(1+z)$: $\mathcal{M}_{c,\rm obs} =
(1+z)\,\mathcal{M}_{c, \rm phys}$. Below, $\mathcal{M}_c$ will
always refer to the observed quantity. To leading order in
amplitude, the GW polarizations are \bea
h_{+}(t)&=&2 \mathcal{M}_c^{5/3} d_L^{-1}(1+\cos^2(\iota))\omega^{2/3}(t_0-t)\,\cos[2\Phi(t_0-t;M,\eta)+\Phi_0],\\
h_{\times}(t)&=&4 \mathcal{M}_c^{5/3}
d_L^{-1}\cos(\iota)\omega^{2/3}(t_0-t)\,\sin[2\Phi(t_0-t;M,\eta)+\Phi_0],
\ena where $\iota$ is the angle of inclination of the binary's
orbital angular momentum with the line-of-sight, $\omega(t_0-t)$
the angular velocity of the equivalent one-body system around the
binary's center-of-mass, and $\Phi(t_0-t;M,\eta)$ the
corresponding orbital phase. The parameters $t_0$ and $\Phi_0$ are
constants giving the epoch of merger and the orbital phase of the
binary at that epoch, respectively. The phase $\Phi$ has been
computed perturbatively in the so-called post-Newtonian formalism
(see \cite{BlanchetLivRevRel} and references therein). Since we
will mostly be concerned with binary neutron stars, spin will not
be important, in which case the phase is known up to 3.5PN in the
usual notation \cite{3.5PN}, and this is what we will use here.

During the inspiral, the change in orbital frequency over a single
period is negligible, and it is possible to apply a stationary
phase approximation to compute the Fourier transform
$\mathcal{H}(f)$ of the time domain waveform $h(t)$. One has \bea
\mathcal{H}(f)=\mathcal{A}f^{-7/6}\exp\left[i(2\pi
ft_0-\pi/4+2\psi(f/2)-\varphi_{(2,0)})\right], \label{waveform}
\ena where the Fourier amplitude $\mathcal{A}$ is given by \bea
\mathcal{A}= \frac{1}{d_L} \sqrt{F_+^2(1+\cos^2(\iota))^2 +
F_\times^2 4 \cos^2(\iota)} \sqrt{\frac{5\pi}{96}}\pi^{-7/6}
\mathcal{M}_c^{5/6}. \ena The functions $\psi$ and
$\varphi_{(2,0)}$ are \bea \psi(f)&=&
-\psi_0+\frac{3}{256\eta}\sum_{i=0}^7\psi_i(2\pi
Mf)^{i/3},\\
\varphi_{(2,0)} &=&
\tan^{-1}\left(-\frac{2\cos(\iota)F_{\times}}{(1+\cos^2(\iota))F_{+}}\right).
\ena The parameters $\psi_i$ can be found in \cite{gwreview}.
$\mathcal{H}(f)$ is taken to be zero outside a certain frequency
range. The upper cutoff frequency is dictated by the last stable
orbit (LSO), which marks the end of the inspiral regime and the
onset of the final merger. We will assume that this occurs when
the radiation frequency reaches $f_{\rm upper}=2f_{LSO}$, with
$f_{LSO}=1/(6^{3/2}2\pi M_{\rm obs})$ the orbital frequency at LSO,
and $M_{\rm obs} = (1+z)\,M_{\rm phys}$ the observed total mass.

In this paper we shall focus on the observation of GW sources by
the Einstein Telescope, a third-generation ground-based
gravitational-wave detector. Although the basic design of ET is
still under discussion, one possibility is to have three
interferometers with $60^\circ$ opening angles and 10km arm
lengths, arranged in an equilateral triangle \cite{et}. The
corresponding antenna pattern functions are: \bea
F_{+}^{(1)}(\theta,\phi,\psi)&=&\frac{\sqrt{3}}{2}\left[\frac{1}{2}(1+\cos^2(\theta))\cos(2\phi)\cos(2\psi)-\cos(\theta)\sin(2\phi)\sin(2\psi)\right],\nonumber\\
F_{\times}^{(1)}(\theta,\phi,\psi)&=&\frac{\sqrt{3}}{2}\left[\frac{1}{2}(1+\cos^2(\theta))\cos(2\phi)\sin(2\psi)+\cos(\theta)\sin(2\phi)\cos(2\psi)\right],\nonumber\\
F_{+,\times}^{(2)}(\theta,\phi,\psi)&=&F_{+,\times}^{(1)}(\theta,\phi+2\pi/3,\psi),\nonumber\\
F_{+,\times}^{(3)}(\theta,\phi,\psi)&=&F_{+,\times}^{(1)}(\theta,\phi+4\pi/3,\psi).
\label{et-F} \ena

The performance of a GW detector is characterized by the one-side
noise {\it power spectral density} $S_h(f)$ (PSD), which plays an
important role in the signal analysis. We take the noise PSD of ET
to be \cite{freise}\footnote{This PSD corresponds to one possible
design of ET; the same reference \cite{freise} also discusses
alternatives.} \bea
S_h(f)=S_0\left[x^{p_1}+a_1x^{p_2}+a_2\frac{1+b_1x+b_2x^2+b_3x^3+b_4x^4+b_5x^5+b_6x^6}{1+c_1x+c_2x^2+c_3x^3+c_4x^4}\right],
\label{et-S} \ena where $x\equiv f/f_0$ with $f_0=200$Hz, and
$S_0=1.449\times 10^{-52}\,{\rm Hz}^{-1}$. The other parameters
are as follows: \bea
p_1=-4.05,&&p_2=-0.69,\nonumber\\
a_1=185.62,&&a_2=232.56,\nonumber\\
b_1=31.18,b_2=-64.72,b_3=52.24,&&b_4=-42.16, b_5=10.17, b_6=11.53\nonumber\\
c_1=13.58,c_2=-36.46,&&c_3=18.56, c_4=27.43. \ena For data
analysis proposes, the noise PSD is assumed to be essentially
infinite below a certain lower cutoff frequency $f_{\rm lower}$
(see the review \cite{gwreview}). For ET we take this to be
$f_{\rm lower}=1~{\rm Hz}$.


The waveforms in Eq.~(\ref{waveform}) depend on the seven free
parameters $(\ln \mathcal{M}_c, \ln
\eta,t_0,\Phi_0,\cos(\iota),\psi,\ln d_L)$; note that for `useful'
events the sky position will be known. In order to deal with the
parameter estimation, throughout this paper, we employ the Fisher
matrix approach \cite{gwfisher}. Comparing with the Markov chain
Monte Carlo (MCMC) analysis, the Fisher information matrix
analysis is simple and accurate enough to estimate the detection
abilities of the future experiments. In the case of a single
interferometer $A$ ($A = 1, 2, 3$), the Fisher matrix is given by
\bea \Lambda^{(A)}_{ij}=\langle \mathcal{H}^{(A)}_i,
\mathcal{H}^{(A)}_j\rangle,~~~\mathcal{H}^{(A)}_i=\partial
\mathcal{H}^{(A)}(f)/\partial p_i, \ena where $\mathcal{H}^{(A)}$
is the output of interferometer $A$, and the $p_i$ denote the free
parameters to be estimated, which are \be
(\ln\mathcal{M}_c,\ln\eta,t_0,\Phi_0,\cos(\iota),\psi,\ln d_L).
\ee The angular brackets denote the scalar product, which, for any
two functions $a(t)$ and $b(t)$ is defined as \bea \langle
a,b\rangle=4\int_{f_{\rm lower}}^{f_{\rm upper}}\frac{\tilde{a}(f)
\tilde{b}^*(f)+\tilde{a}^*(f)\tilde{b}(f)}{2}\frac{df}{S_h(f)},
\label{innerproduct} \ena where $\tilde{a}$ and $\tilde{b}$ are
the Fourier transforms of the functions $a(t)$ and $b(t)$. The
Fisher matrix for the combination of the three independent
interferometers is then \be \Lambda_{ij} = \sum_{A=1}^3
\Lambda^{(A)}_{ij}. \ee

The inner product also allows us to write the signal-to-noise
ratios $\rho^{(A)}$, $A=1,2,3$ in a compact way: \be \rho^{(A)} =
\sqrt{\langle \mathcal{H}^{(A)}, \mathcal{H}^{(A)} \rangle}. \ee
The combined signal-to-noise ratio for the network of the three
independent interferometers is then \be \rho = \left[ \sum_{A =
1}^3 \left(\rho^{(A)}\right)^2 \right]^{1/2}. \ee

In this paper, we shall focus on the estimation of the parameter
$\ln d_L$. The 1-$\sigma$ observational error, $\sigma_o$, can be
estimated from the Fisher matrix $\Lambda_{ij}$. An important
point is that shGRBs are believed to be beamed: the $\gamma$
radiation is emitted in a narrow cone more or less perpendicular
to the plane of the inspiral. We will take the total beaming angle
to be at most 40$^\circ$ \cite{grb} (corresponding to $\iota \leq
20^\circ$). It will be assumed that shGRBs are produced by the
mergers of the neutron star binaries. For definiteness we take
them to have component masses of $(1.4,1.4)M_{\rm sun}$. We will
consider $1000$ sources up to a redshift of $z=2$, which is where
the angle-averaged signal-to-noise ratio approximately reaches the
value 8 for sources with $\iota < 20^\circ$.

Before continuing, we mention that coalescing binaries composed of
a neutron star and a black hole (NSBH) could also cause shGRBs
\cite{grb}. For a fixed distance, a NSBH event will have a larger
SNR than a BNS event, leading to an improved measurement of $\ln
d_L$. The intrinsic event rates for NSBH are quite uncertain, but
they are expected to be considerably lower than for BNS
\cite{ratespaper}; on the other hand, NSBH events will be visible
to ET out to redshifts of $z \sim 4$ \cite{sath}. It is likely
that the inclusion of NSBH would have a noticeable beneficial
effect on the determination of cosmological parameters, especially
if the black holes have spin, but this we leave for future
studies.

For a given event, distance measurements will be subject to two
kinds of uncertainties: the instrumental error $\sigma_o$ which
can be estimated using a Fisher matrix as discussed above, and an
additional error $\sigma_l$ due to the effects of weak lensing. As
in previous work \cite{sath} we assume the contribution to the
distance error from weak lensing to satisfy $\sigma_l=0.05z$.
Thus, the total uncertainty on $\Delta \ln d_L$ is taken to be
 \bea
 \Delta \ln d_L=\sqrt{\sigma_o^2+\sigma_l^2}. \label{delta-dl}
 \ena

In the next subsection we discuss how the information from
multiple GW standard sirens can be combined to compute the
expected measurement uncertainties on cosmological parameters.

\subsection{Gravitational-wave standard sirens\label{section2c}}

Now let us turn to the determination of the cosmological
parameters, including dark energy parameters, by the GW standard
sirens. For each shGRB source, the luminosity distance $d_L$ is
measured from the GW observation, and the redshift $z$
can be obtained from the electromagnetic counterpart. Thus,
the $d_L-z$ relation can be employed to constrain various
cosmological parameters. For the cosmological model introduced in
Sec.~\ref{section2a}, we consider five free parameters
$(w_0,w_a,\Omega_m,\Omega_k,h_0)$ which can be constrained by
GW standard sirens. We note that the value of
$\Omega_{de}$ (the relative energy density of the dark energy
component) is determined by the quantities $\Omega_m$ and
$\Omega_k$ through $\Omega_{de}\equiv 1-\Omega_{m}-\Omega_{k}$.

In order to estimate the errors on these parameters, we study a
Fisher matrix $F^{\rm GW}_{ij}$ for a collection of inspiral
events: \bea F^{\rm GW}_{ij}=\sum_{k}\frac{\partial_{i}(\ln
d_{L}(k))\partial_{j}(\ln d_{L}(k))}{(\Delta \ln
d_L(k))^2},\label{fisher0}\ena where the indices $i$ and $j$ run
from 1 to 5, denoting the free parameters
$(w_0,w_a,\Omega_m,\Omega_k,h_0)$. Eq.~(\ref{dl}) gives the
expression for $d_L$, and the partial derivatives with respect to
the parameters are evaluated at the parameter values corresponding
to the fiducial cosmological model of Sec.~\ref{section2a}. The
uncertainty $\Delta \ln d_L$ is calculated by using Eq.
(\ref{delta-dl}). The index $k=1,2,\ldots$, labels the event at
($z_k,\hat{\gamma}_k$), where the vector $\hat{\gamma}$ stands for
the angles $(\theta,\phi,\iota,\psi)$. Here we should mention
that, in (\ref{fisher0}), we have ignored the photometric redshift
errors and the possible errors generated by the peculiar
velocities of the sources relative to the Hubble flow
\cite{kocsis}. Given that the majority of our sources will be at
$z > 0.4$, we do not expect the latter to make much difference to
our main results.

\begin{figure}
\begin{center}
\includegraphics[width=14cm,height=10cm]{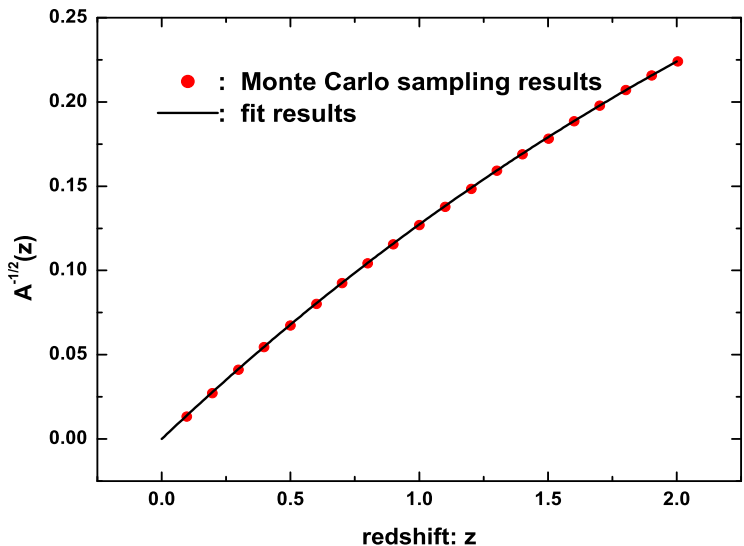}
\end{center}\caption{ The averaged quantity $A^{-1/2}(z)$ [defined in (\ref{average})] as a function of redshift $z$. The red dots denote the results based on
the Monte Carlo sampling, and the solid line denotes the fit
results.}\label{figure0}
\end{figure}

Since $d_L$ is independent of $\hat{\gamma}$, this Fisher matrix
can be rewritten as \bea F^{\rm
GW}_{ij}=\sum_{z_k}\partial_{i}(\ln d_{L}(z_k))\partial_{j}(\ln
d_{L}(z_k)) \left\{\sum_{\hat{\gamma}_k}\frac{1}{(\Delta \ln
d_L(z_k,\hat{\gamma}_k))^2}\right\}.\label{fisher1}\ena When the
number of events is large, the sum over events in (\ref{fisher1}) can
be replaced by an integral, so that we obtain \bea F^{\rm
GW}_{ij}={\int_0^2
\partial_{i}(\ln d_{L})\partial_{j}(\ln d_{L})f(z)A(z)dz},
\label{fisher2} \ena where $f(z)$ is the number distribution of
the GW sources over redshift $z$. $A(z)$ is the average of
${1}/{(\Delta \ln d_L(z,\hat{\gamma}))^2}$ over the angles
$(\theta,\phi,\iota,\psi)$ with the constraint $\iota < 20^\circ$:
 \bea
 A(z)\equiv\left\langle\frac{1}{(\Delta \ln
d_L(z,\hat{\gamma}))^2}\right\rangle_{\hat{\gamma};\,\,\,\iota < 20^\circ}.
\label{average}
 \ena
In order to calculate the averaged quantity $A(z)$, we used a
Monte Carlo sampling with 10,000 choices of $\hat{\gamma}$
for a given $z$, where $z$ ranges from 0 to 2 in steps of 0.1.
The results are indicated in Fig. \ref{figure0} by the red dots. We find that
these points can be fit very accurately by a simple relation
(see the black solid line in Fig.~\ref{figure0}):
 \bea
 A^{-1/2}(z)= 0.1449z-0.0118z^2+0.0012z^3 \label{fit},
 \ena
which is used in our subsequent calculation.

In (\ref{fisher2}), the upper integration limit $z=2$ is the
redshift at which the angle-averaged signal-to-noise ratio is
approximately 8 \cite{sath}. In our fiducial model with
$\Omega_{k}=0$, the number distribution $f(z)$ is given by \bea
f(z)=\frac{4\pi \mathcal{N}r(z)d^2_C(z)}{H(z)(1+z)}, \ena where
$d_{C}$ is the comoving distance, which is defined as
$d_{C}(z)\equiv \int_{0}^{z}1/H(z')dz'$. The function $r(z)$
describes the time evolution of the burst rate, and the constant
$\mathcal{N}$ (the number of the sources per comoving volume at
redshift $z=0$ over the observation period) is fixed by requiring
the total number of the sources $N_{\rm GW}= \int_0^2f(z)dz$. The
expected total number of inspirals per year within the horizon of
ET is $\sim$ several $\times 10^{5}$ for neutron star binaries.
If, as suspected, neutron star binaries are progenitors of shGRBs
\cite{grb}, it might be possible to make a coincident detection of
a significant subset of the events in the GW and electromagnetic
windows, which can then be considered as standard sirens. As we
have mentioned, shGRBs are believed to be beamed with small
beaming angle, so only a small fraction of the total number of
neutron star binaries are expected to be observed as shGRBs.
Following \cite{sath}, we assume that about $1000$ events
($\sim10^{-3}$ of the total number of binary coalescences) will be
observed in both windows, i.e., $N_{\rm GW}=1000$ throughout this
paper.

\begin{figure}
\begin{center}
\includegraphics[width=14cm,height=10cm]{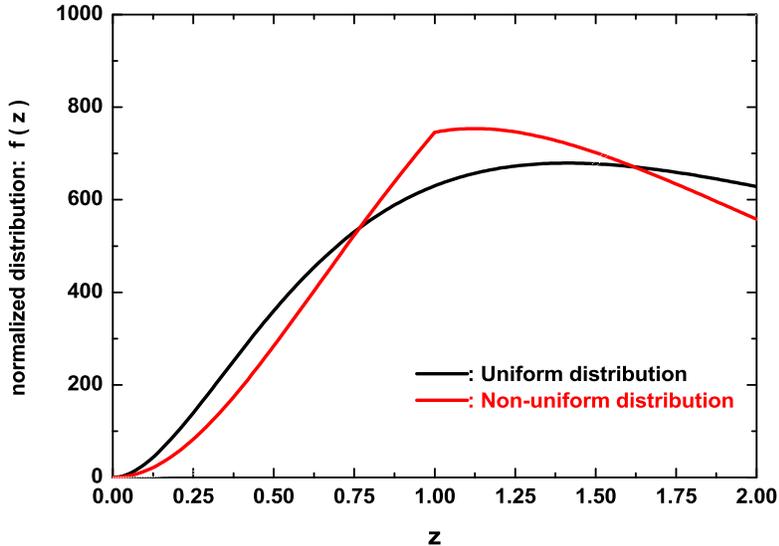}
\end{center}\caption{ The normalized distribution of the GW sources.}\label{figure2}
\end{figure}

Since the time evolution of the source rate is as yet unclear, in
this paper we shall consider two different forms for the function
$r(z)$. In the first case we assume that the sources are
distributed uniformly, i.e., with constant comoving number density
throughout the redshift range $0\le z \le 2$ (hereafter we will
refer to this as the uniform distribution). In this case we have
$r(z)=1$, which is what was assumed in the previous work
\cite{sath}. In the other case, we take $r(z)$ to be the following
function: $r(z)=(1+2z)$ for $z\le 1$, $r(z)=(15-3z)/4$ for
$1<z<5$, and $z=0$ for $z\ge 5$. This approximate fit to the rate
evolution is suggested in \cite{nonuniform}. Hereafter, we shall
call this the nonuniform distribution. In Fig. \ref{figure2}, we
plot the distribution function $f$ as a function of redshift $z$
in the two cases. Note that in the case with nonuniform
distribution, the sources are a little more concentrated at $z=1$.
In what follows we will find out how this affects the
uncertainties on the cosmological parameters.

Using the definition (\ref{fisher2}), we can calculate the Fisher
matrix $F^{\rm GW}_{ij}$ for the following two cases : {\it a. GW
events with uniform distribution; b. GW events with nonuniform
distribution.} The results are listed in Tables I and II.

\begin{table}
\caption{GW Fisher matrix in the case with uniform distribution}
\begin{center}
\label{table2}
\begin{tabular}{|c|c|c|c|c|c|}
    \hline
     & $w_0$ & $w_a$ & $\Omega_m$ & $\Omega_k$ & $h_0$  \\
    \hline
    $w_0$& $0.273681\times10^{4}$ & $0.433681\times10^{3}$ & $0.753068\times10^{4}$  & $0.261216\times10^{4}$& $0.221885\times10^{5}$ \\
   \hline
   $w_a$& $0.433681\times10^{3}$ & $0.753710\times10^{2}$ & $0.128638\times10^{4}$  & $0.406152\times10^{3}$& $0.317462\times10^{4}$ \\
  \hline
     $\Omega_m$ & $0.753068\times10^{4}$ & $0.128638\times10^{4}$ & $0.221212\times10^{5}$  & $0.704345\times10^{4}$& $0.571447\times10^{5}$ \\
   \hline
    $\Omega_k$& $0.261216\times10^{4}$ & $0.406152\times10^{3}$ & $0.704345\times10^{4}$  & $0.251558\times10^{4}$& $0.213013\times10^{5}$ \\
   \hline
    $h_0$& $0.221885\times10^{5}$ & $0.317462\times10^{4}$ & $0.571447\times10^{5}$  & $0.213013\times10^{5}$& $0.216280\times10^{6}$ \\
  \hline
\end{tabular}
\end{center}
\end{table}
\begin{table}
\caption{GW Fisher matrix in the case with nonuniform
distribution}
\begin{center}
\label{table4}
\begin{tabular}{|c|c|c|c|c|c|}
    \hline
     & $w_0$ & $w_a$ & $\Omega_m$ & $\Omega_k$ & $h_0$  \\
    \hline
    $w_0$& $0.256794\times10^{4}$ & $0.427648\times10^{3}$ & $0.731269\times10^{4}$  & $0.244368\times10^{4}$& $0.194634\times10^{5}$ \\
   \hline
   $w_a$& $0.427648\times10^{3}$ & $0.762633\times10^{2}$ & $0.129200\times10^{4}$  & $0.399934\times10^{3}$& $0.303753\times10^{4}$ \\
  \hline
     $\Omega_m$ & $0.731269\times10^{4}$ & $0.129200\times10^{4}$ & $0.219941\times10^{5}$  & $0.682599\times10^{4}$& $0.529628\times10^{5}$ \\
   \hline
    $\Omega_k$& $0.244368\times10^{4}$ & $0.399934\times10^{3}$ & $0.682599\times10^{4}$  & $0.234666\times10^{4}$& $0.186267\times10^{5}$ \\
   \hline
    $h_0$& $0.194634\times10^{5}$ & $0.303753\times10^{4}$ & $0.529628\times10^{5}$  & $0.186267\times10^{5}$& $0.162814\times10^{6}$ \\
  \hline
\end{tabular}
\end{center}
\end{table}

\begin{figure}
\begin{center}
\includegraphics[width=18cm,height=8cm]{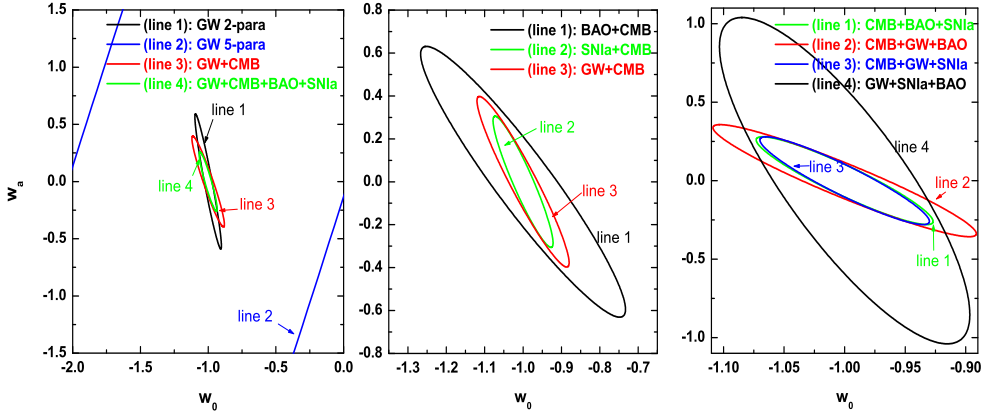}
\end{center}\caption{The
two-dimensional uncertainty contours of the dark energy parameters
$w_0$ and $w_a$ in the case with uniform
distribution.}\label{figure3}
\end{figure}

From the Fisher matrices, we can calculate the 1-$\sigma$
uncertainties on the parameters, which are $\Delta
p_i=\sqrt{(F^{{\rm GW}})^{-1}_{~~ii}}$. For the case with uniform
distribution, by using the results in Table I, we find that
 \bea
 \Delta w_0=2.62,~\Delta
 w_a=9.53,~\Delta\Omega_m=0.815,~\Delta\Omega_{k}=2.03,~\Delta
 h_0=1.20\times10^{-2}. \label{gw-5p}
 \ena
We plot the two-dimensional uncertainty contour of the parameters
$w_0$ and $w_a$ in Fig. \ref{figure3} (blue curve, i.e., line 2,
in the left panel). Unfortunately, we find that the error bars on
the parameters are all fairly large, especially for the dark
energy parameters $w_0$ and $w_a$. This is caused by the strong
degeneracy between ($w_0$, $w_a$) and the other parameters
($\Omega_m$, $\Omega_k$, $h_0$), especially ($\Omega_m$,
$\Omega_k$). To illustrate this, let us do the following
calculation. First we fix the values of the parameters
($\Omega_m$, $\Omega_k$, $h_0$) to be their fiducial values, and
only consider ($w_0$, $w_a$) as free parameters. By using the
results in Table I, we obtain
 \bea
 \Delta w_0=0.064,~\Delta
 w_a=0.388. \label{gw-2p}
 \ena
We find that the values of $\Delta w_0$ and $\Delta w_a$ become
much smaller in this case. The two-dimensional uncertainty contour
of $w_0$ and $w_a$ is also plotted in Fig. \ref{figure3} (black
curve, i.e. line 1, in the left panel). This figure shows that
there is correlation between the parameters $w_0$ and $w_a$.
Recall that a goal of the dark energy programs is to test whether
dark energy arises from a simple cosmological constant, ($w_0=-1$,
$w_a=0$). For a given data set we can do better (as far as
excluding the cosmological constant model is concerned) than
simply quoting the values of $\Delta w_0$ and $\Delta w_a$. This
is because the effect of dark energy is generally not best
constrained at $z=0$. For the phenomenological form of the EOS of
the dark energy $w(z)=w_0+w_az/(1+z)$, the constraint on $w(z)$
varies with the redshift $z$. So, similar to \cite{detf}, we can
define the best pivot redshift, denoted as $z_p$, where the
uncertainty in $w(z)$ equals the uncertainty in a model that
assumes $w_a=0$. In this paper, we denote the EOS at this best
pivot redshift as $w_p\equiv w(z_p)$. The best pivot redshift
$z_p$ can be calculated by $z_p=-1/(1+\frac{\Delta w_a}{\rho\Delta
w_0})$, where $\rho$ is the correlation coefficient of $w_0$ and
$w_a$. The value of $\Delta w_p$ is calculated by $\Delta
w_p=\Delta w_0\sqrt{1-\rho^2}$. In this case (two free
parameters), the results for $z_p$ and $\Delta w_p$ are \bea
z_p=0.188,~~\Delta w_p=0.019. \label{gw-2p-best} \ena The value of
$\Delta w_p$ as well as that of $\Delta w_a$ are commonly used to
describe the detection ability of the experiments \cite{detf}.

On the other hand, we can also fix the values of the parameters
($w_0$, $w_a$) to be their fiducial values, and only consider
($\Omega_m$, $\Omega_k$, $h_0$) as free parameters. By using the
results in Table I, we obtain
 \bea
 \Delta\Omega_m=0.021,~\Delta\Omega_{k}=0.087,~\Delta
 h_0=5.48\times10^{-3}.
 \ena
Again we find that the values of these errors, especially the
values of $\Delta \Omega_m$ and $\Delta \Omega_k$, are much
smaller that those in Eq. (\ref{gw-5p}). These results show that
the GW standard sirens can constrain the dark energy parameters
rather well, on condition that we can break the strong degeneracy
between the parameters ($w_0$, $w_a$) and the parameters
($\Omega_m$, $\Omega_k$, $h_0$). In the next subsection, we will
find that this can be realized if we consider the CMB observations
as a prior.

\begin{figure}
\begin{center}
\includegraphics[width=16cm,height=16cm]{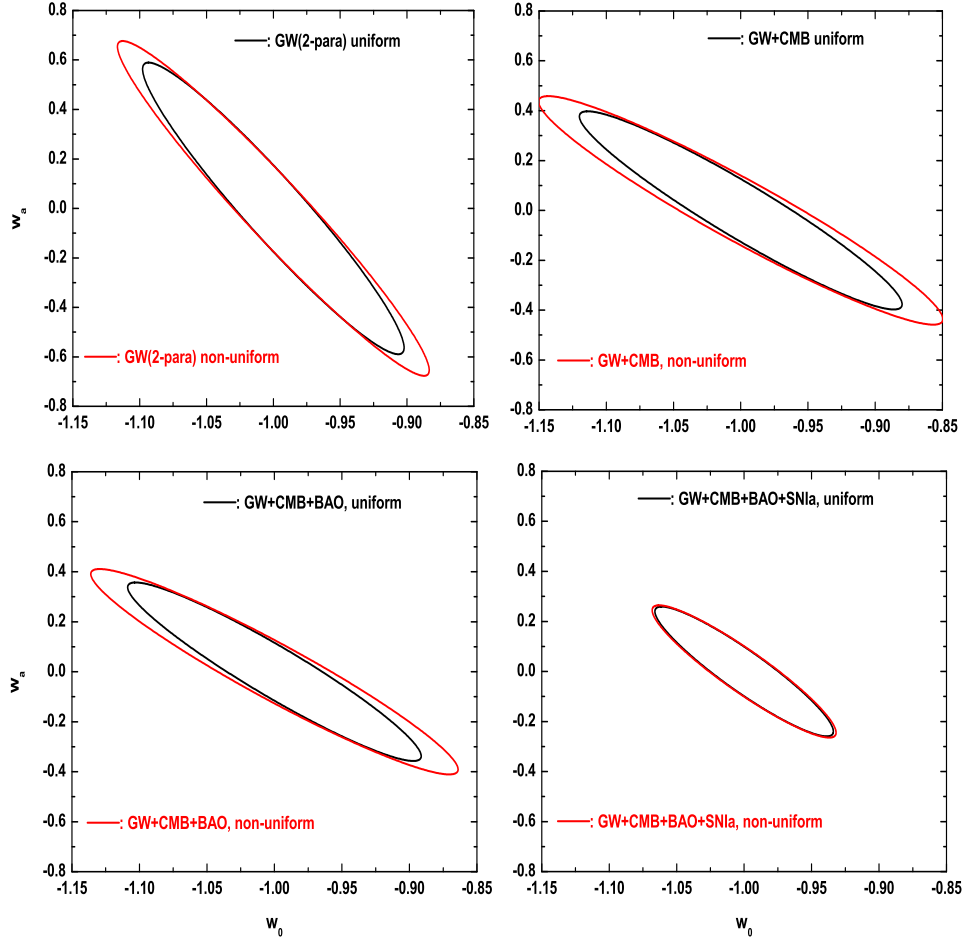}
\end{center}\caption{The two-dimensional uncertainty contours for the dark energy parameters $w_0$ and $w_a$, in the different cases.}\label{figure4}
\end{figure}

To conclude this subsection we discuss the determination of the
dark energy parameters ($w_0$, $w_a$) by ET observations in two
cases considered in this paper. By using the results in Table I
and II, we calculate the errors of the two parameters (the other
parameters are fixed at their fiducial values). The
two-dimensional uncertainty contours are shown in Fig.
\ref{figure4} (upper left panel). This figure shows that the
errors of the parameters are a little larger for the nonuniform
distribution than those in the corresponding case with the uniform
distribution. This is because the sources in the nonuniform
distribution are a little more concentrated at the redshift $z=1$.
Hence the number of the sources in the high redshift and the low
redshift regions is smaller, making it more difficult to constrain
the dark energy's evolution. So, in addition to the number of the
sources, the redshift distribution of the sources also plays a
crucial role for the detection of dark energy. The results listed
in (\ref{gw-2p}) constitute the optimistic case among the cases we
have considered. On the other hand, it is helpful to list the
results in the case with nonuniform distribution, which are
 \bea
 \Delta w_0=0.077,~\Delta
 w_a=0.445. \label{gw-2p-2}
 \ena
We find that these uncertainties are a little larger than those in
(\ref{gw-2p}). We can also calculate the values of $z_p$ and
$\Delta w_p$ for this case, which are \bea z_p=0.200,~~\Delta
w_p=0.020. \label{gw-2p-2-best} \ena The value of $\Delta w_p$ is
also a little larger than that in the case with uniform
distribution.

\subsection{{Planck} CMB prior \label{section2d}}

As will be clear from the discussion above, the ability of GW
sources to constrain dark energy depends strongly on how well the
background parameters $\Omega_m$ and $\Omega_k$ can be measured
beforehand. Indeed, if one tries to determine these background
parameters as well as the dark energy parameters together using GW
sources, the values of $\Delta w_0$ and $\Delta w_a$ become very
large, and the constraints on dark energy become meaningless.
Hence we should consider another observational method which can
determine these background parameters through a prior observation.
As we shall see, this is also necessary for the other ways used to
study dark energy -- BAO and SNIa -- so that it does not diminish
the value of having self-calibrating standard sirens: GW
observations will provide us with an important independent check.

Observations of the Cosmic Microwave Background (CMB) temperature and polarization anisotropies are always used
as the required prior. The WMAP satellite has already given fairly good results for the CMB $TT$ (temperature-temperature
auto-correlation) and $TE$ (temperature-polarization cross
correlation) power spectra in the multipole range $\ell<800$. The
background parameters $(\Omega_m,\Omega_k)$ have already been well determined by
7-year WMAP observations; for example, the constraint on the curvature is
$-0.0133<\Omega_k<0.0084$ ($95\%$ C.L.) \cite{wmap}. These
constraints are expected to be greatly improved by the {Planck}
observations \cite{planck}, which will give good data
on the CMB $TT$ and $TE$ power spectra up to
$\ell\sim2000$. In addition, {Planck} is also expected to
observe the $EE$ (electric type polarization) power spectrum. In
this subsection, we shall consider the potential CMB observations
by the {Planck} satellite. For the noise power spectra of {
Planck}, we consider the instrumental noises at three
frequency channels: at $100$GHz, $143$GHz, and $217$GHz. For the
reduced foreground radiations (including dust and synchrotron), we
assume that the reduced factor $\sigma^{\rm fg}=0.1$, i.e. $10\%$
residual foregrounds are considered as the noises. The total noise
power spectra $N_{\ell}^{TT}$ and $N_{\ell}^{EE}$ of the Planck
satellite can be found in \cite{zhao}. Note that in this paper we
assume 4 sky (28 months) survey.

In order to estimate the constraints on the cosmological
parameters, we shall again use the Fisher information matrix
technique. The Fisher matrix is calculated by \cite{cmbfisher}
\begin{eqnarray}
\label{fishermatrix}
 F^{\rm CMB}_{{i}{j}}=\sum_{\ell=2}^{\ell_{\rm max}}\sum_{XX',YY'}\frac{\partial {C}_{\ell}^{XX'}}
{\partial p_{i}}{{\rm Cov}^{-1} (D_{\ell}^{XX'},D_{\ell}^{YY'})}
\frac{\partial {C}_{\ell}^{YY'}}{\partial p_{j}},
\end{eqnarray}
where $p_i$ are the cosmological parameters to be evaluated.
$C_{\ell}^{XX'}$ are the CMB power spectra and
$D_{\ell}^{XX'}$ are their estimates. ${\rm Cov}^{-1}$ is the
inverse of the covariance matrix. The non-vanishing components of
the covariance matrix are given by
 \begin{eqnarray}
 {\rm Cov} (D_{\ell}^{XX},D_{\ell}^{XX})&=&\frac{2}{(2\ell+1)f_{\rm sky}}
(C_{\ell}^{XX}+N_{\ell}^{XX})^2~~~(X=T,E),\nonumber\\
 {\rm Cov} (D_{\ell}^{TE},D_{\ell}^{TE})&=&\frac{1}{(2\ell+1)f_{\rm sky}}
[(C_{\ell}^{TE})^2+(C_{\ell}^{TT}+N_{\ell}^{TT})(C_{\ell}^{EE}+N_{\ell}^{EE})],
\nonumber\\
 {\rm Cov} (D_{\ell}^{TT},D_{\ell}^{EE})&=&\frac{2}{(2\ell+1)f_{\rm sky}}
(C_{\ell}^{TE})^2,\nonumber\\
 {\rm Cov} (D_{\ell}^{TT},D_{\ell}^{TE})&=&\frac{2}{(2\ell+1)f_{\rm sky}}
C_{\ell}^{TE}(C_{\ell}^{TT}+N_{\ell}^{TT}),\nonumber\\
 {\rm Cov} (D_{\ell}^{EE},D_{\ell}^{TE})&=&\frac{2}{(2\ell+1)f_{\rm sky}}
C_{\ell}^{TE}(C_{\ell}^{EE}+N_{\ell}^{EE}).\nonumber
 \end{eqnarray}
Note that in the calculation we have adopted $\ell_{\rm max}=2000$,
and the sky-cut factor $f_{\rm sky}=0.65$ suggested by Planck
Bluebook \cite{planck}.

The CMB power spectra $C_{\ell}^{XX'}$ depend on all the
cosmological parameters, including the background parameters and
the perturbation parameters. In the calculation, we first build
the Fisher matrix for the full nine parameters
$(w_0,w_a,\Omega_bh_0^2,\Omega_ch_0^2,\Omega_k, n_s, A_s, h_0,
\tau)$. In order to obtain the constraint on the parameter
$\Omega_m$, we change the full Fisher matrix to the new one of the
nine parameters $(w_0,w_a,\Omega_m,\Omega_ch_0^2,\Omega_k, n_s,
A_s, h_0, \tau)$, where
$\Omega_m=(\Omega_bh_0^2+\Omega_ch_0^2)/h_0^2$ has been used. In
order to directly compare and combine with the Fisher matrix of GW
method, we marginalize the new nine-parameter Fisher matrix to the
one with five parameters $(w_0,w_a,\Omega_m,\Omega_k,h_0)$. The
results are shown in Table III. The errors of the parameters are
given by $\Delta p_i=\sqrt{(F^{\rm CMB})^{-1}_{~~~ii}}$. By using
the Fisher matrix in Table III, we obtain
 \bea
 \Delta w_0=0.411,~\Delta
 w_a=0.517,~\Delta\Omega_m=8.88\times10^{-2},~\Delta\Omega_{k}=2.27\times10^{-3},~\Delta
 h_0=0.115.
 \ena
This result shows that {Planck} alone can give quite tight
constraints on $\Omega_m$ and $\Omega_k$, which is just
complementary with the GW method. However, CMB observations alone
cannot constrain the parameters $w_0$ and $w_a$,
which is because that the CMB power spectra are only sensitive to
the physics in the early Universe at $z\sim 1100$, where dark
energy is totally subdominant.

\begin{table}
\caption{CMB Fisher matrix}
\begin{center}
\label{CMBtable1}
\begin{tabular}{|c|c|c|c|c|c|}
    \hline
     & $w_0$ & $w_a$ & $\Omega_m$ & $\Omega_k$ & $h_0$  \\
    \hline
    $w_0$& $0.414303\times10^{5}$ & $0.115085\times10^{5}$ & $0.287229\times10^{6}$  & $-0.678690\times10^{6}$& $0.339923\times10^{6}$ \\
   \hline
   $w_a$&$0.115085\times10^{5}$  &$0.320204\times10^{4}$  & $0.797415\times10^{5}$ &$-0.190373\times10^{6}$ &$0.943498\times10^{5}$  \\
  \hline
     $\Omega_m$ & $0.287229\times10^{6}$ & $0.797415\times10^{5}$ & $0.219854\times10^{7}$& $-0.465663\times10^{7}$ &$0.251813\times10^{7}$  \\
   \hline
    $\Omega_k$& $-0.678690\times10^{6}$& $-0.190373\times10^{6}$ & $-0.465663\times10^{7}$ & $0.136912\times10^{8}$& $-0.548091\times10^{7}$  \\
   \hline
    $h_0$& $0.339923\times10^{6}$& $0.943498\times10^{5}$& $0.251813\times10^{7}$& $-0.548091\times10^{7}$ &$0.291587\times10^{7}$   \\
  \hline
\end{tabular}
\end{center}
\end{table}

We now investigate the combination of CMB and GW
methods. In order to do this, we define a new Fisher matrix, which
is the sum of $F_{ij}^{\rm GW}$ and $F_{ij}^{\rm CMB}$. By using this new
Fisher matrix, we obtain
 \bea
 \Delta w_0=0.079,~\Delta
 w_a=0.261,~\Delta\Omega_m=5.14\times10^{-3},~\Delta\Omega_{k}=6.66\times10^{-4},~\Delta
 h_0=5.96\times10^{-3},
 \ena
for the uniform distribution case. We find that the values of
$\Delta w_0$ and $\Delta w_a$ are fairly close to those in Eq.
(\ref{gw-2p}), where we have only considered the GW observations
but assumed the backgrounds parameters $\Omega_m$, $\Omega_k$ and
$h_0$ are fixed. These results show that taking the CMB
observation as a prior is nearly equivalent to fixing the
parameters $\Omega_m$, $\Omega_k$ and $h_0$. The two-dimensional
uncertainty contour of the parameters $w_0$ and $w_a$ is shown in
Fig. \ref{figure3} (red lines, i.e. line 3, in the left panel and
the middle panel). We can also calculate the best pivot redshift
$z_p$ and the value of $\Delta w_p$, which are \bea
z_p=0.401,~~\Delta w_p=0.025. \ena

By the same method, we also obtain the results in the nonuniform
distribution case. The two-dimensional uncertainty contour of
$w_0$ and $w_a$ is shown in Fig. \ref{figure4} (upper-right
panel). In this case, the results are
 \bea
 \Delta w_0=0.099,~\Delta
 w_a=0.302,~\Delta\Omega_m=7.30\times10^{-3},~\Delta\Omega_{k}=6.70\times10^{-4},~\Delta
 h_0=8.99\times10^{-3}. \label{gw-2-cmb}
 \ena
The best pivot redshift $z_p$ and the uncertainty $\Delta w_p$ are
\bea z_p=0.454,~~~\Delta w_p=0.030. \ena Again as expected, we
find that the values are larger than the corresponding values in
the uniform distribution case.



\section {Detection of dark energy by BAO and SNIa observations, and the comparison with ET-GW observations \label{section3}}

In the above we found that by combining the potential {Planck} CMB
observation with ET-GW observations, one can get fairly tight
constraints on the dark energy parameters $w_0$ and $w_a$. In this
section we discuss the detection abilities of the other two
probes: BAO and SNIa. Currently these two methods play the crucial
role for the determination of the dark energy component. In the
near future, the detection abilities of these two methods are
expected to be significantly improved; in this section a detailed
discussion is given. All three probes, BAO, SNIa and GW, constrain
the EOS of dark energy by probing the large-scale background
geometry of the Universe (different from the weak gravitational
lensing method \cite{weaklensing}), so a fair comparison can be
made, as we shall do here.

\subsection{Detection of dark energy by potential BAO observations \label{section3a}}

{The BAO method relies on the distribution of
baryonic matter to infer the redshift-distance relation. The
characteristic scale length of structure which can be accurately
determined from the CMB is used as a standard rod. By measuring
the angular size of this characteristic scale-length as a function
of redshift, the effect of dark energy can be inferred.} The BAO
method can constrain the dark energy by two observable quantities
$\ln\left(H(z)\right)$ and $\ln\left(d_A(z)\right)$, where $H(z)$
is the binned Hubble parameter and $d_A(z)$ is comoving angular
diameter distance, which is related to the luminosity distance by
$d_{A}=d_{L}/(1+z)$. Similar to the quantity $d_L$, these two observables only depend on the cosmological
parameters $w_0$, $w_a$, $\Omega_m$, $\Omega_k$ and $h_0$, which will be considered as the parameters determined by the observations.

In order to investigate the constraints on the cosmological
parameters, we build the following Fisher information matrix \cite{detf}:
\bea F_{ij}^{\rm BAO}=\sum_{k}\frac{\partial
\ln\left(H(z_k)\right)}{\partial p_i}\frac{\partial
\ln\left(H(z_k)\right)}{\partial
p_j}\left(\frac{1}{\sigma_{\ln\left(H(z_i)\right)}}\right)^2
+\frac{\partial \ln\left(d_A(z_k)\right)}{\partial
p_i}\frac{\partial \ln\left(d_A(z_k)\right)}{\partial
p_j}\left(\frac{1}{\sigma_{\ln\left(d_A(z_i)\right)}}\right)^2.
\ena
Again the index $k$ denotes the observables, which are binned into several redshift bins. $p_i$ denotes the cosmological parameters.
$\sigma_{\ln\left(H(z)\right)}$ and $\sigma_{\ln\left(d_A(z)\right)}$ are the errors (including the observational errors and the systematic errors)
of the observables $\ln\left(H(z)\right)$ and $\ln\left(d_A(z)\right)$, respectively.
In order to study the detection ability of the BAO method, we shall consider the potential observations of a typical project, the final JDEM (Joint Dark Energy Mission)
project \cite{detf}, which is expected to survey $10000$ deg$^2$ in
the redshift range $z\in(0.5,2)$. In the calculation, we bin the observables $\ln\left(H(z)\right)$ and $\ln\left(d_A(z)\right)$ into
$10$ redshift bins, i.e. $\Delta z=0.15$ for each bin. The calculation of the theoretical values of these quantities are straightforward.
For the errors of these observable data, we use the fitting formulae derived in \cite{bao-error} (see also Eq. (4.8) in \cite{detf}).

The results of the Fisher information matrix are shown in Table
IV. To begin with we consider a simple case with only two free
parameters ($w_0$, $w_a$). We assume that the other parameters
($\Omega_m$, $\Omega_k$, $h_0$) are fixed to their fiducial
values. By using the Fisher matrix in Table IV, we obtain the
uncertainties of the free parameters: \bea \Delta
w_0=0.087,~~\Delta w_a=0.346,~~ z_p=0.323, ~~\Delta w_p=0.023.
\label{bao-2p} \ena

However, if we try to constrain all five parameters by BAO
observations, the uncertainties of the parameters will become
fairly large. For instance, the uncertainties of $w_0$ and $w_a$
become $\Delta w_0=0.850$ and $\Delta w_a=3.611$, respectively, which are much
larger than those in (\ref{bao-2p}). Similarly to the discussion in
Sec. \ref{section2d}, we can consider the combination of the BAO
observation and the Planck CMB prior. By analogous steps we
obtain the results
 \bea
 \Delta w_0=0.176,~\Delta
 w_a=0.415,~\Delta\Omega_m=2.01\times10^{-2},~\Delta\Omega_{k}=6.40\times10^{-4},~\Delta
 h_0=2.57\times10^{-2}. \label{bao-cmb}
 \ena
The best pivot redshift and the value of $\Delta w_p$ are
\bea
z_p=0.664,~~\Delta w_p=0.059.
\ena

\begin{table}
\caption{BAO Fisher matrix}
\begin{center}
\label{BAOtable1}
\begin{tabular}{|c|c|c|c|c|c|}
    \hline
     & $w_0$ & $w_a$ & $\Omega_m$ & $\Omega_k$ & $h_0$  \\
    \hline
    $w_0$& $0.193253\times10^{4}$ & $0.471352\times10^{3}$ & $0.865506\times10^{4}$  & $0.288000\times10^{4}$& $0.125880\times10^{5}$ \\
   \hline
   $w_a$& $0.471352\times10^{3}$ & $0.123299\times10^{3}$ & $0.234243\times10^{4}$  & $0.788613\times10^{3}$& $0.309758\times10^{4}$ \\
  \hline
     $\Omega_m$ & $0.865506\times10^{4}$ & $0.234243\times10^{4}$ & $0.457760\times10^{5}$  & $0.153244\times10^{5}$& $0.579934\times10^{5}$ \\
   \hline
    $\Omega_k$& $0.288000\times10^{4}$ & $0.788613\times10^{3}$ & $0.153244\times10^{5}$  & $0.541597\times10^{4}$& $0.190452\times10^{5}$ \\
   \hline
    $h_0$& $0.125880\times10^{5}$ & $0.309758\times10^{4}$ & $0.579934\times10^{5}$  & $0.190452\times10^{5}$& $0.834616\times10^{5}$ \\
  \hline
\end{tabular}
\end{center}
\end{table}

\subsection{Detection of dark energy by potential SNIa observations\label{section3b}}

Now, let us turn to discuss the detection of dark energy
parameters by the SNIa probe. Type Ia supernovae serve as a
standard candle of (approximately) known luminosity. The redshift
of supernova can be obtained by studying its spectral lines. Thus
the redshift-distance relation can be gotten from SNIa surveys.
Now, SNIa observed from various experiments have been used
successfully to deduce the acceleration of the Universe after
$z=1$ \cite{SNIa,SNIa-others}. In the near future, observations of
SNIa are expected to be significantly improved, so that they will
continue to serve as one of the most important methods for the
determination of dark energy.

The observables for SNIa data are the apparent magnitudes $m$,
which can be corrected to behave as standard candles with absolute
magnitude $M$ with $m=M+\mu(z)$. The function $\mu(z)$ for the
measured redshift is \bea \mu(z)=5\log_{10}(d_L(z))+25, \ena where
$d_{L}$ is the luminosity distance. In this paper, we shall
consider SNIa observations by the future SNAP
(Supernova/Acceleration Probe) project \cite{snap}. As suggested
by the SNAP group, we consider $300$ low redshift supernovae,
uniformly distributed over $z\in(0.03,0.08)$. The error bar on the
magnitude is assumed to be $\sigma_m=0.15$ mag. In addition,
$2000$ high redshift supernovae in the range $z\in(0.1,1.7)$ are
considered. The expected redshift distribution of these sources
can be found in the SNAP white book (the middle red curve in Fig.
9 of \cite{snap}). We bin these 2000 sources into 10 redshift bins
in the range $z\in(0.1,1.7)$. The total errors of the observables
$\sigma(z)$ can be estimated as follows \cite{snap,snap2}:
 \bea
\sigma=\sqrt{\sigma_1^2+\sigma_2^2}, \label{snia-error}\ena where
$\sigma_1=0.15\,{\rm mag}/\sqrt{N}$ ($N$ is the total number of
supernovae in each bin) is the intrinsic random Gaussian error,
and $\sigma_2=0.02{\rm mag}(1+z)/2.7$ is the error due to the
astrophysical systematics.

Thus, we can build a Fisher information matrix, which is \bea
F_{ij}^{\rm SN}=\sum_{k}\frac{\partial \mu(z_k)}{\partial
p_i}\frac{\partial \mu(z_k)}{\partial
p_j}\left(\frac{1}{\sigma(z_k)}\right)^2 . \ena The results are
shown in Table V. The errors of the cosmological parameters are
estimated by $\Delta p_i=\sqrt{(F^{\rm SN})^{-1}_{~~~ii}}$.
Similar to Sec. \ref{section3a}, we first discuss the simplest
case with two free parameters ($w_0$, $w_a$). The results are \bea
\Delta w_0=0.054, ~~\Delta w_a=0.302,~~z_p=0.211, ~~\Delta
w_p=0.012. \label{SNIa-2p} \ena To constrain all five cosmological
parameters, we consider the combination of SNIa and the Planck CMB
prior to decouple the degeneracy between ($w_0$, $w_a$) and the
other parameters. The errors of the parameters are
 \bea
 \Delta w_0=0.051,~\Delta
 w_a=0.201,~\Delta\Omega_m=3.49\times10^{-3},~\Delta\Omega_{k}=6.52\times10^{-4},~\Delta
 h_0=3.39\times10^{-3}. \label{SNIa-cmb}
 \ena
We find that the values of $\Delta w_0$ and $\Delta w_a$ in
(\ref{SNIa-cmb}) are close to those in (\ref{SNIa-2p}). The best
pivot redshift and the value of $\Delta w_p$ are also obtained
\bea
z_p=0.313,~~\Delta w_p=0.019.
\ena

\begin{table}
\caption{SNIa Fisher matrix}
\begin{center}
\label{SNIatable1}
\begin{tabular}{|c|c|c|c|c|c|}
    \hline
     & $w_0$ & $w_a$ & $\Omega_m$ & $\Omega_k$ & $h_0$  \\
    \hline
    $w_0$& $0.703955\times10^{4}$ & $0.122773\times10^{4}$ & $0.207309\times10^{5}$  & $0.669019\times10^{4}$& $0.522067\times10^{5}$ \\
   \hline
   $w_a$& $0.122773\times10^{4}$ & $0.225085\times10^{3}$ & $0.377761\times10^{4}$  & $0.115185\times10^{4}$& $0.849589\times10^{4}$ \\
  \hline
     $\Omega_m$ & $0.207309\times10^{5}$ & $0.377761\times10^{4}$ & $0.636028\times10^{5}$  & $0.194262\times10^{5}$& $0.146902\times10^{6}$ \\
   \hline
    $\Omega_k$& $0.669019\times10^{4}$ & $0.115185\times10^{4}$ & $0.194262\times10^{5}$  & $0.639862\times10^{4}$& $0.498026\times10^{5}$ \\
   \hline
    $h_0$& $0.522067\times10^{5}$ & $0.849589\times10^{4}$ & $0.146902\times10^{6}$  & $0.498026\times10^{5}$& $0.491151\times10^{6}$ \\
  \hline
\end{tabular}
\end{center}
\end{table}

\subsection{Comparison with the ET-GW observations}

Now let us compare the detection abilities of various probes: GW,
BAO and SNIa. First we consider the simplest case, where only dark
energy parameters ($w_0$, $w_a$) are considered. The errors of the
parameters are given in (\ref{bao-2p}) for BAO, and in
(\ref{SNIa-2p}) for SNIa. We find that the values of $\Delta w_0$,
$\Delta w_a$ and $\Delta w_p$ are all smaller for the SNIa probe.
This shows that, comparing with the JDEM BAO project, the SNAP
SNIa project is expected to give a tighter constraint on the dark
energy. For the ET-GW project, in Sec. \ref{section2c}, we have
considered two cases. For the uniform distribution case, the
results are given in (\ref{gw-2p}) and (\ref{gw-2p-best}). On the
other hand, if the nonuniform distribution case is considered, the
results are shown in (\ref{gw-2p-2}) and (\ref{gw-2p-2-best}).
Comparing with these results, we conclude that, in both cases, the
detection ability of ET-GW method is stronger than that of the
JDEM BAO project, but weaker than that of the SNAP SNIa project.
This is mainly because the number of the GW standard sirens ($\sim
1000$ as we have assumed) is smaller than that of the SNIa
standard candles ($\sim 2000$ at high redshift). In order to
clearly show this, let us consider another case for ET-GW method,
where we assume that $2000$ sources with nonuniform distribution
will be observed. We then obtain $\Delta w_0=0.054$, $\Delta
w_a=0.315$ and $\Delta w_p=0.014$, which comes close to the
projected uncertainties of the SNAP SNIa project given by
(\ref{SNIa-2p}). We note that the relative error bars of the SNIa
in Eq. (51) are larger than those of GW sources in (32),
especially at low redshifts. However, we find that this
disadvantage of the SNIa method is overcome by the assumed 300 low
redshift SNIa in $z\in (0.03, 0.08)$.

We can also compare the results of these three probes, when
considering the full 5 cosmological parameters and adopting the
Planck CMB prior. In Fig. \ref{figure3} (middle panel), we plot
the two-dimensional uncertainty contours of the parameters ($w_0$,
$w_a$), where for the ET-GW method we have considered the case
with uniform distribution. Similarly, we find that the red curve
(GW+CMB) is only a little looser than the green one (SNIa+CMB),
but much tighter than the black one (BAO+CMB). In Fig.
\ref{figure3} (right panel), we plot the results of the
two-dimensional uncertainty contours for the four combinations
(CMB+BAO+SNIa, CMB+GW+BAO, CMB+GW+SNIa, GW+SNIa+BAO). We find that
the first three combinations have similar results. However, the
constraint on the dark energy parameters is much looser for
combination of GW+SNIa+BAO, where the Planck CMB probe is absent.
This panel shows that the CMB prior indeed plays a crucial role in
the detection of dark energy.

Let us now combine all the four probes to constrain the
cosmological parameters, including the dark energy parameters. If
we consider the uniform distribution for the GW sources, we obtain
the constraints on the dark energy parameters
 \bea
 \Delta w_0=0.044,~\Delta w_a=0.171,~z_p=0.308,~\Delta w_p=0.017. \label{four-1}
 \ena
This is the best constraint what we could expect to obtain. If we
consider the case with nonuniform distribution for the GW sources,
the constraints slightly loosen to
 \bea
 \Delta w_0=0.045,~\Delta w_a=0.174,~z_p=0.313,~\Delta w_p=0.017. \label{four-2}
 \ena
In Fig. \ref{figure4} (lower-right panel), we plot the
two-dimensional uncertainty contours of the parameters ($w_0$,
$w_a$) for both cases. We find that the two curves are very close
to each other; the relative weight of the GW probe is not very
high for the combined methods.

\begin{figure}
\begin{center}
\includegraphics[width=15cm,height=12cm]{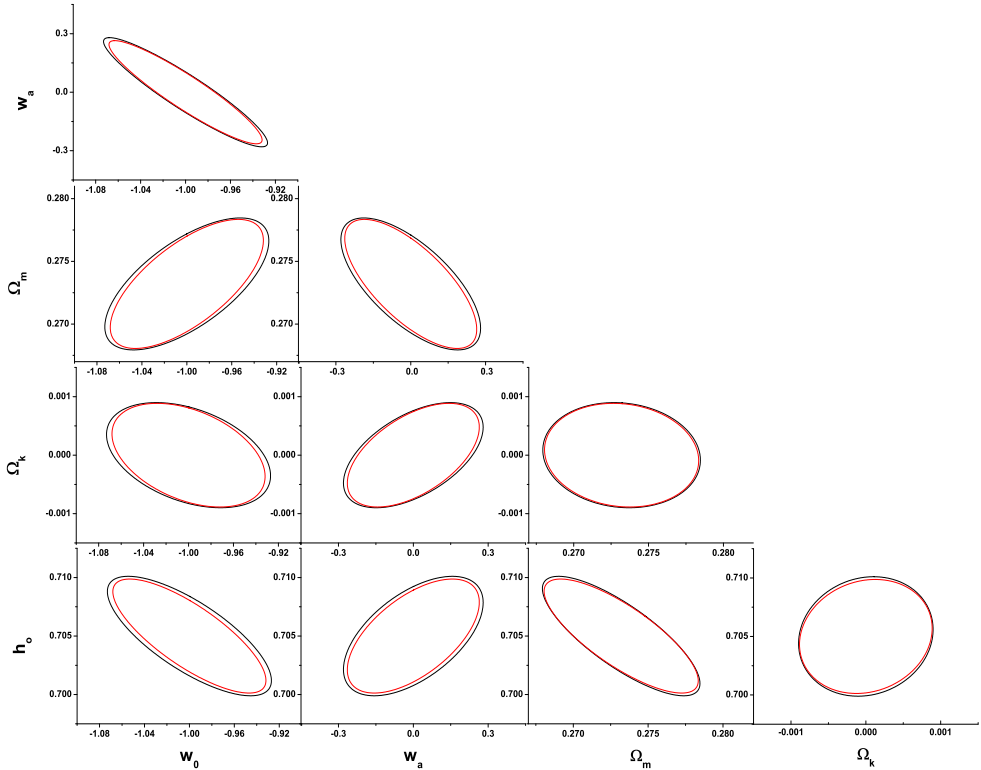}
\end{center}\caption{ The two-dimensional uncertainty contours of the cosmological parameters. The
black (larger) curves denote the results of CMB+BAO+SNIa, and the
red (smaller) curves denote the results of CMB+BAO+SNIa+GW. Here,
we have assumed the GW with nonuniform distribution (The case with
uniform distribution gives the nearly overlapped results
).}\label{figure5}
\end{figure}

Finally, we would like to know how much the ET-GW probe can
contribute in constraining all the five cosmological parameters
($w_0$, $w_a$, $\Omega_m$, $\Omega_k$, $h_0$). In order to do so,
we first calculate the constraints on the parameters by the
combination of CMB+BAO+SNIa. We obtain \bea \Delta
w_0=0.048,~\Delta w_a=0.184,~ \Delta\Omega_m=3.46\times
10^{-3},~\Delta\Omega_k=5.91\times 10^{-4},\Delta h_0=3.36\times
10^{-3}.
\ena If we then add the contribution of the ET-GW probe with
nonuniform distribution (the uniform distribution case gives the
very close results), the results become \bea \Delta w_0=0.045,
~\Delta w_a=0.174,~ \Delta\Omega_m=3.39\times
10^{-3},~\Delta\Omega_k=5.83\times 10^{-4},\Delta h_0=3.20\times
10^{-3}. \ena In this case we find that, due to the contribution
of ET-GW probe, $\Delta w_0$ is decreased by a $6.3\%$ and $\Delta
w_a$ by $5.5\%$. In Fig. \ref{figure5} we plot the two-dimensional
uncertainty contours of the free parameters ($w_0$, $w_a$,
$\Omega_m$, $\Omega_k$, $h_0$) for both combinations.

\section{Conclusion \label{section4}}

If short-hard $\gamma$-ray bursts (shGRBs) are produced by the
mergers of neutron star binaries, the luminosity distances $d_L$
of the sources can be determined by the Einstein Telescope
gravitational-wave detector in the redshift range $z < 2$. The
redshifts $z$ of the sources can be determined with great accuracy
through their electromagnetic counterparts. Thus it will be
possible to use shGRBs as `standard sirens' to study the dark
energy component in the Universe by determining the EOS and its
time evolution.

When calculating the uncertainties in the determination of $d_L$
by the ET observations, we assumed that the $\gamma$-ray emission
is confined to a cone with an opening angle as large as $40^\circ$
\cite{grb}, corresponding to inclination angles $\iota <
20^\circ$. In order to study the effect of the redshift
distribution of the sources, we considered two different kinds of
distributions: one in which sources are distributed uniformly in
comoving volume, and a nonuniform distribution as in
\cite{nonuniform}. We found that, by taking into account the
Planck CMB prior, the errors on the dark energy parameters are
expected to be $\Delta w_0=0.079$ and $\Delta w_a=0.261$ in the
uniform distribution case, which is close to the detection ability
of the SNAP Type Ia Supernovae project. Even in the `pessimistic'
case with nonuniform distribution of the sources, the errors are
$\Delta w_0=0.099$ and $\Delta w_a=0.302$, which is weaker than
the detection ability of the SNAP Type Ia Supernova project, but
stronger than that of the JDEM Baryon Acoustic Oscillation
project. We also found that, comparing with the combination of the
future CMB(Planck)+BAO(JDEM)+SNIa(SNAP) projects, the contribution
of this kind of standard sirens can decrease the error of $w_0$ by
$\sim 6.3\%$ and that of $w_a$ by $\sim 5.5\%$. Thus, the kind of
self-calibrating GW standard sirens accessible to Einstein
Telescope would provide an excellent probe of the dark energy
component.

Finally, it is important to mention that, in addition to GW, CMB,
BAO and SNIa methods, there are a number of other probes,
including cosmic weak lensing, galaxy clustering, and so on, which
can also be used to detect the dark energy component in the
Universe (see \cite{detf,wang} for details). In practice, one
should combine all the probes. However, in this paper, we have
emphasized that using shGRBs as `standard sirens' constitutes an
important complement to the general electromagnetic methods.


\section*{Acknowledgements}

The authors thank L.~P.~Grishchuk and B. S.~Sathyaprakash for
stimulating discussions. W.Z. is partially supported by Chinese
NSF Grants Nos. 10703005, 10775119 and 11075141. C.V.D.B. and
T.G.F.L. are supported by the research programme of the Foundation
for Fundamental Research on Matter (FOM), which is partially
supported by the Netherlands Organisation for Scientific Research
(NWO).


.







\end{document}